\documentclass[aps,pre,superscriptaddress,showpacs,twocolumn]{revtex4}
\usepackage{graphicx, amssymb, amsmath}
\begin{document}
\title{Decoherence and single electron charging in an electronic Mach-Zehnder interferometer}
\author{L. V. Litvin }
\altaffiliation[Permanent address:] { Institute of Semiconductor
Physics, 630090, Novosibirsk, Russia}
\author{H.-P. Tranitz }
\author{W. Wegscheider }
\author{C. Strunk}
\affiliation{ \mbox{Institut f\"{u}r experimentelle und angewandte
Physik, Universit\"{a}t  Regensburg, D-93040 Regensburg, Germany}}
%
%
%
\begin{abstract}
We investigate the temperature and voltage dependence of the quantum
interference in an electronic Mach-Zehnder interferometer using edge
channels in the integer quantum-Hall-regime. The amplitude of the
interference fringes is significantly smaller than expected from
theory; nevertheless the functional dependence of the visibility on
temperature and bias voltage agrees very well with theoretical
predictions. Superimposed on the Aharonov-Bohm (AB) oscillations, a
conductance oscillation with six times smaller period is observed.
The latter depends only on gate voltage and not on the AB-phase, and
may be related to single electron charging.
\end{abstract}
%
%
\maketitle

Electron interferences in mesoscopic conductors mani\-fest itself in
conductance oscillations which are $h/e$ periodic in the magnetic
field $B$ \cite{webb_review}. In conventional metals the visibility
$\nu_I$ of these Aharonov-Bohm (AB) oscillations typically amounts
to $\nu_I=G_0/G\simeq 10^{-3}$, where $G$ is the conductance of the
sample and $G_0=2e^2/h$ the conductance quantum. The visibility can
be enhanced to $\nu_I\simeq 0.1$ by reducing the number of
conductance channels, e.g. in nanostructures based on
two-dimensional electron systems \cite{olsh,hansen} containing only
a small number of conductance channels.  Recently an electronic
analogue \cite{seelig} of the well known optical Mach-Zehnder
interferometer (MZI) was realized, which employs the one-dimensional
edge channels in the integer quantum Hall regime \cite{heiblum1}. In
these devices single channel interference can be realized, while
backscattering processes are suppressed. This results in measured
visibilities up to $\nu_I\simeq 0.6$. Such devices open a path for
the realization of fundamental two-particle interference experiments
in the spirit of Hanbury-Brown and Twiss \cite{HBT}, as recently
proposed \cite{sukh}.

We have realized MZIs similar to those in
Ref.~\onlinecite{heiblum1}. A quantum point contact (QPC) is used to
partition an edge channel leaving contact S into two paths. After
propagation of the edge channel along the inner and outer edge of an
ring-shaped mesa, the two paths are brought to interference at a
second QPC, resulting in two output channels D1 and D2 of the
interferometer (see Fig.~1). The phase of the two partial waves can
be changed both by magnetic field and by an electrostatic gate G.
The measured visibility is much smaller than expected from theory
\cite{chung} and reported in Ref.~\onlinecite{heiblum1}. Despite
this quantitative disagreement, the functional dependence of $\nu_I$
on $T$ and $V_{bias}$ fits very well the simple model of
Ref.~\onlinecite{chung}. In addition to the AB-oscillations we found
another type of conductance oscillations, which have a significantly
smaller period in gate voltage when compared to the AB period.

The mesa was prepared through wet etching of a modulation doped
GaAs/Ga$_{x}$Al$_{1-x}$As heterostructure containing a
two-dimensional electron system (2DES) 90~nm below the surface.
At~4~K, the unpatterned 2DES density and mobility were
$n$=2.0$\times$10$^{15}$~m$^{-2}$ and $\mu$=206~m$^{2}$/(V$\,$s),
respectively. Using standard electron beam litho\-graphy techniques,
we prepared split gates connected by air bridges, defining QPCs of
500~nm length and 220~nm gap width. The measurements were performed
in a dilution refrigerator with two stages of copper powder filters
at bath temperature and at 100~mK. The currents at contacts D1 and
D2 were measured using a lock-in technique with an ac excitation
voltage (1~kHz) of 10$\div$16~$\mu$V applied to contact S. We
checked that the measured visibilities remained constant below this
level of the voltage.

%
\begin{figure}
\includegraphics[width=55mm]{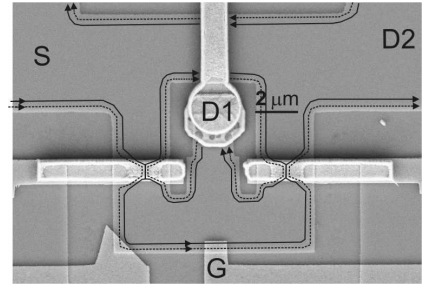}
\caption{SEM image of Mach-Zehnder interferometer with the scheme of
edge states for filling factor 2. The inner edge channel (full line)
is split into two branches using QPCs.}
\end{figure}

To characterize the samples, we first verified that the current
through contact D1 goes to zero when the filling factor $\nu$
approaches 1 or 2 (see Fig.~2).  This implies a complete suppression
of backscattering between the edge channels at the opposite edges of
the mesa (Fig.2, curve D1). The sum of both detector currents,
$I_{D1}+I_{D2}$ shows quantized current (conductance) levels which
allows us to determine the value of $\nu$. From the position of the
minima in current D1 (marked by vertical arrows in Fig.2) at
magnetic fields $B_{\nu}=\frac{n\,h}{\nu\,e}$ for $\nu=3,4,6$ we
determine an electron density $n$=1.7$\times$10$^{15}$~m$^{-2}$
which is 15~\% smaller than that of the unpatterned 2DES because of
the depletion near the mesa edges. If we apply a negative voltage at
one of the QPCs at $\nu\approx2$, the inner edge channel starts to
split into two branches flowing into D1 and D2. When the QPC is
pinched off the current will be completely redirected from D2 into
D1. Then, the conductance $G_{S,D1}$ will be close to 2$e^{2}/h$
(inset to Fig.2). In our experiment the conductance $G_{S,D1}$ at
the plateau reaches about 95\% of ($e^{2}/h$). This implies a high
transmission of the miniature ohmic contact D1, reflecting only 5\%
of the incident current back into the edge channel.

To observe interference both QPCs were tuned to transmission $1/2$
for the inner edge channel (full line at Fig.1). The currents at D1
and D2 as a function of gate voltage $V_G$ are shown at Fig.3(a),
3(b) respectively. Current conservation requires that the
oscillations detected in D1 and D2 should go in antiphase and sum up
to a constant value [Fig.3(c)]. Two periods of oscillation are seen
in these traces. The amplitude of the small period oscillations is
not constant and shows a beating-like pattern. Fourier analysis
reveals two periods of 0.71~and 0.80~mV for the small period
oscillations and a period of 5.1~mV for the large period
oscillations [Fig.3(d)].
\begin{figure}
\includegraphics[width=85mm]{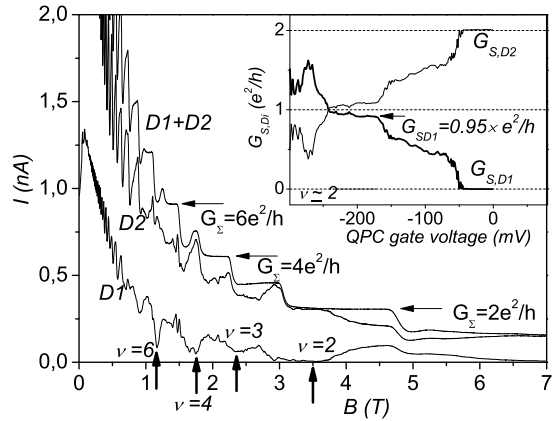}
\caption{The currents collected by D1, D2 and their sum as function
of magnetic field at $V_{\mathrm{bias}}$ = 4~$\mu$V and $T$=25~mK,
when both QPCs are open. The thick vertical arrows show magnetic
fields corresponding to the indicated filling factors; the
horizontal arrows label conductance level.  Inset: two point
conductance between contacts S, D1 and S,D2 as function of  gate
voltage at one of the QPCs for $B=3.26$~T; the second QPC is open.}
\end{figure}
%
\begin{figure}
\includegraphics[width=75mm]{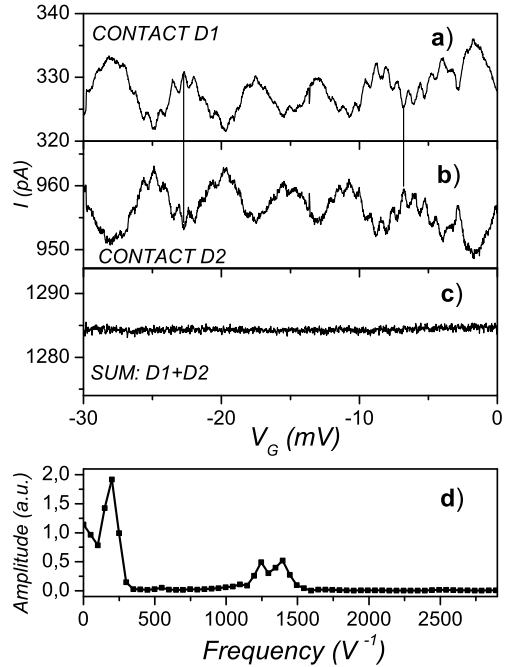}
\caption{The currents collected by D1 (a), D2 (b), their sum (c) as
function of gate G voltage at $T$=25~mK, $B$=3.27~T ($\nu$=2.14).
(d) Typical Fourier spectrum of oscillations.}
\end{figure}
Since the AB-period $h/e$ estimated from the area enclosed by the
edge channels amounts to $~78\;\mu$T, a direct measurement of the
magnetic field dependence of the interference pattern requires
control of the magnetic field at a level of $10^{-6}$. Since this is
difficult to achieve, we have exploited the gradual decay of the
magnetic field in persistent mode at a rate of about 20~$\mu$T/h. In
this way the time delay between successive traces can be translated
into changes of $B$. In Fig.~4 we show a sequence of gate sweeps
recorded with time delays of 10 min. The large period oscillations
shift linearly with time delay to the left, indicating that these
oscillations are of the expected Aharonov-Bohm type. Besides the
regular oscillations we observed an occasional switching of the
phase of the large period oscillation by $\simeq\pi$ (see, e.g.,
dashed line in Fig.~4). We attribute these events to slow random
transitions of charged impurities between two metastable states. The
maximum visibility for the observed AB~oscillations was 1.5\%. In
contrast to these, the small period oscillations do not shift with
time and thus do not depend on magnetic field (Fig.4). This fact
points towards an electrostatic origin of this effect.
\begin{figure}[t]
\includegraphics[width=75mm]{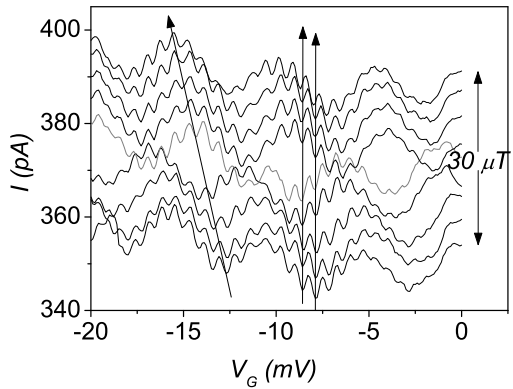}
\caption{The time evolution of oscillation at $T$=25~mK,
$B$=3.27~T~($\nu$=2.14) indicated by the gradual decay of magnetic
field in persistant mode. Successive traces are recorded with 10 min
time delay (traces are vertically shifted for clarity by 5~pA). The
arrow indicate the progressing recording time and the time shift of
the oscillations. The gray curve in the middle displays a phase
lapse by $\pi$ at $V_{G}$=-11~mV and separates two trains with
linear phase evolution.}
\end{figure}

We now discuss the decay of the oscillation amplitude with
temperature and bias voltage. The temperature dependence of the
amplitude for both types of oscillations is displayed in Fig.~5. The
amplitude of the AB~oscillation was averaged first over a few
oscillations in a single trace and than over four sweeps taken at
each temperature. The amplitude of the small period oscillations was
calculated by integrating the Fourier spectrum within a region
including both high frequency peaks, i.e. from 1200~to
1450~$V^{-1}$. From Fig.~5 we can infer a characteristic energy
$E_C\simeq 9\;\mu$eV (corresponding to $T_C$=100~mK) related to the
small period oscillation. The peak at higher frequency
(1400~$V^{-1}$) decays more rapid than the peak at 1250~$V^{-1}$
both with increasing temperature (inset in Fig.~5) and bias voltage.

According to Ref.~\onlinecite{chung} the visibility of the AB
oscillations $\nu_{I}~=~(I_{max}-I_{min})/(I_{max}+I_{min})$ decays
with the temperature or voltage as
\begin{equation}
\nu_{I}=\frac{1}{2}\;\frac{4\pi
k_{B}T}{eV_{bias}}\left[\sinh\left(\frac{k_{B}T\pi}{E_{b}}\right)\right]^{-1}\left|\sin\left(\frac{eV_{bias}}{2E_{b}}\right)\right|.
\end{equation}
\\
The prefactor $1/2$ corresponds to QPC transmission probabilities
$\mathcal{T}_{A}=\mathcal{T}_{B}=1/2$. The $T$- and
$V_{bias}$-dependence scales with another characteristic energy
$E_b$.
 $E_{b}$ determines the phase an electron acquires
traversing the asymmetric interferometer with a difference $\Delta
L$ in path length between the two interferometer arms. An electron
at energy $E$ above the Fermi energy $E_{F}$ collects a phase
difference $\Delta\phi(E)=\Delta \phi(E_{F})+E/E_{b}$, where
$E_{b}\approx \hbar \upsilon_{D}/(\Delta L)$ and $\upsilon_{D}$
being the drift velocity. The energy dependence of this phase
difference smears the interference pattern when the interfering
electrons are spread over energy ranges $k_BT$ and $eV_{bias}$. This
is in complete analogy to the interference of a polychromatic light
beam in optics. Equation~1 well reproduces the shape of the measured
curves (Fig.6) but not the absolute value of the visibility. The
latter is 80~times smaller in our experiment.
%

\begin{figure}
\includegraphics[width=75mm]{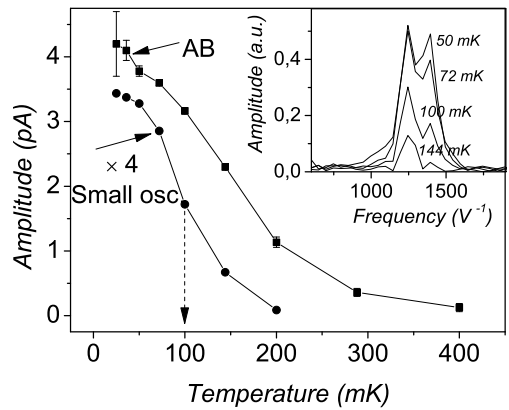}
\caption{Measured amplitude of Aharonov-Bohm (squares) and small
period oscillations (open circles) as function of temperature.
Inset: Evolution of the spectrum of the small oscillations with
temperature.}
\end{figure}

The values of $E_{b}\simeq 18\,\mu$eV extracted from the fit of the
$T$- and $V_{bias}$-dependence of Eq.~1 to the data coincide within
3\% accuracy. The theory assumes a dc voltage bias, while we used an
ac bias in the experiment. Averaging over the ac bias window results
in the solid line in Fig.~6(b), which is again in complete agreement
with the data.

Next we connect the energy scales $E_C$ and $E_b$ with the geometric
parameters of our device. From the location of the inner edge
channel we can deduce the path difference $\Delta L$ between the two
interferometer arms, which determines $E_b$. Taking into account the
depletion region at the mesa edge we estimate the distance between
the mesa edge and the location of inner edge channel. Our estimate
is based on the electron density extracted from the positions of
current minima at integer filling factor in D1 (see Fig.2) for the
MZIs with the arm width of 2.5 and 1.7~$\mu$m. Using the edge
reconstruction model \cite{chklovsk}, we find a depletion length of
$2l$=180~nm, a distance between the mesa edge and the center of the
incompressible strip $x_{1}$=150~nm (for $\nu$=2) and the width of
incompressible strip $a_{1}$=45~nm. This results in a distance
$l+x_{1}+a_{1}/2=260$~nm between the mesa edge and the inner edge
strip for our interferometer, implying that the length of the two
interferometer arms differs by about $\Delta L$=2.0~$\mu$m. The
known $E_b$ and $\Delta L$ yield the drift velocity of
5$\times10^{4}$~m/s which agrees well with other estimates
\cite{komiyama} for the edge state regime.

The question remains, why the measured visibility of the
AB~oscillations for our interferometer is so small. It can be seen
from Fig.~6 that $\nu_I$ remains temperature dependent down to
50~mK. Although a reduction of the ac bias voltage did not improve
the visibility, we can not exclude other sources of an electron
heating. On the other hand an intrinsic source of dephasing may
result from the internal degrees of freedom of the QPC
\cite{cronenwett}, which sensitively depend of the shape of the QPC
potential. In our case the QPCs consist of rather long (500~nm) and
narrow (200~nm) channels, which are quite different from an ideal
saddle point potential.

In earlier experiments, a very high interference visibility of 60\%
was reported \cite{heiblum1,heiblum2}. This value rapidly dropped
down to 1\% when increasing the temperature from 20 to 100~mK,
indicating a relevant energy scale $E_b$ even smaller than in our
experiment. A satisfactory agreement between these experiments with
simple theoretical models \cite{chung} is still lacking, since the
data of Ref.~\onlinecite{heiblum2} indicate a surprising
independence of the visibility on the asymmetry between the
interferometer arms.

\begin{figure}[t]
\includegraphics[width=75mm]{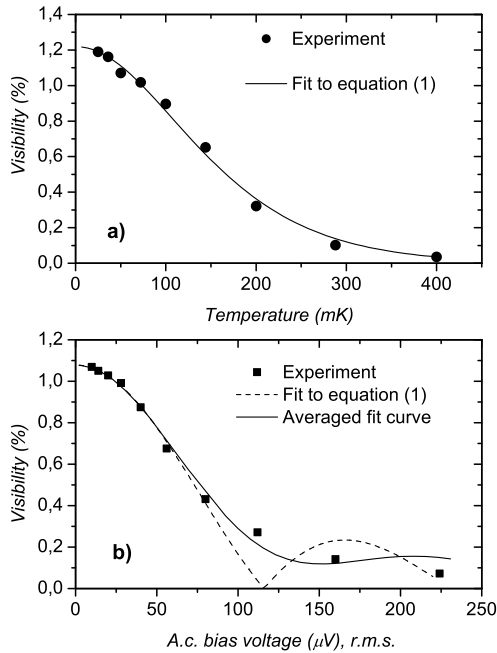}
\caption{Visibility decay as function of (a) temperature ($V_{bias}$
= 16~$\mu$V, r.m.s.) and (b) voltage ($T=25$~mK). In (b) fitting was
carried out in the $V_{bias}$-range 0$\div$80~$\mu$V).}
\end{figure}

We now return to the small oscillations and their characteristic
energy $E_C$. The striking periodicity of these oscillations may
suggest Coulomb blockade as their possible origin. This requires a
charging object with a capacitive coupling to the gate which is
stronger than that of the inner edge channel. The latter is
responsible for the AB interference. The outer edge channel is
ballistically transmitted and should not show charging. Since the
side gate is expected to shift both edge channels without
qualitative changes in their structure, we see no obvious candidate
for such a charging object, unless a relatively large puddle of
electrons is formed underneath the gate. We do not have independent
evidence that this may happen. On the other hand, the splitting of
the observed frequency may suggest that this oscillations related to
the presence of two edge channels in the structure.
%

In conclusion, we investigated single channel interference in
electronic Mach-Zehnder interferometers and found a strong and
unexpected decoherence. Although the absolute value of the measured
visibility is a factor~80 smaller than expected, the functional form
of its suppression by finite temperature, bias voltage and asymmetry
of two interfering paths agrees well with a simple model. In
addition, we found a second type of oscillations which appears to be
of electrostatic origin. This effect requires further investigation.

We thank, M.~Heiblum, J.~Weis, F.~Marquardt, S.~Ludwig and
J.~Kotthaus for stimulating discussions. This work was supported by
the Deutsche Forschungsgemeinschaft in the framework of SFB631
"Solid state quantum information processing".

\end{document}